\documentclass{bmcart}

\usepackage{graphicx}
\RequirePackage{natbib}
\usepackage[utf8]{inputenc} 

\startlocaldefs
\endlocaldefs

\begin{document}

\begin{frontmatter}

\begin{fmbox}


\title{Using word embeddings to analyse audience effects and individual differences in parenting Subreddits}


\author[
   addressref={aff1, aff2},                   
   email={melody.sepahpourfard@ul.ie}   
]{\inits{MS}\fnm{Melody} \snm{Sepahpour-Fard}}
\author[
   addressref={aff3,aff4,aff5},
   email={Mike.Quayle@ul.ie}
]{\inits{MQ}\fnm{Michael} \snm{Quayle}}
\author[
   addressref={aff6},
   email={schuld@ukzn.ac.za}
]{\inits{MS}\fnm{Maria} \snm{Schuld}}
\author[
   addressref={aff7,aff8},
   corref={aff7,aff8}, 
   email={taha.yasseri@ucd.ie}
]{\inits{TY}\fnm{Taha} \snm{Yasseri}}


\address[id=aff1]{
  \orgname{Science Foundation Ireland Centre for Research Training in Foundations of Data Science}, 
  \cny{Ireland}                                    
}
\address[id=aff2]{%
  \orgname{Department of Mathematics and Statistics, University of Limerick},
  \street{D\"{u}sternbrooker Weg 20},
  \postcode{24105}
  \city{Castletroy, Limerick},
  \cny{Ireland}
}
\address[id=aff3]{
  \orgname{Centre for Social Issues Research, University of Limerick}, 
  \city{Castletroy, Limerick},              
  \cny{Ireland}                                    
}
\address[id=aff4]{
  \orgname{Department of Psychology, University of Limerick}, 
  \city{Castletroy, Limerick},              
  \cny{Ireland}                                    
}
\address[id=aff5]{
  \orgname{Department of Psychology, School of Applied Human Sciences, University of KwaZulu-Natal}, 
  \city{Durban, KwaZulu-Natal},                 
  \cny{South Africa}                            
}
\address[id=aff6]{
  \orgname{Department of Psychology, University of Johannesburg}, 
  \city{Johannesburg},                 
  \cny{South Africa}                           
}
\address[id=aff7]{
  \orgname{School of Sociology, University College Dublin}, 
  \city{Dublin},                     
  \cny{Ireland}                          
}
  \address[id=aff8]{
  \orgname{Geary Institute for Public Policy, University College Dublin}, 
  \city{Dublin},                     
  \cny{Ireland}  
}



\end{fmbox}


\begin{abstractbox}

\begin{abstract} 

In this paper, we explore the language of gender-specific groups of "mothers" and "fathers" in comparison to the same individuals interacting just as "parents." Human beings adapt their language to the audience they interact with. While audience effects have been studied theoretically and in small-scale research designs, large-scale studies of naturally-occurring audience effects are rare. To study the impact of audience and gender in a natural setting, we choose a domain where gender plays a particularly salient role: parenting. We look at the audience effects in interaction with gendered contexts emphasizing different social identities (i.e., mother, father, and parent). As a natural setting, we choose to study interactions on Reddit. We collect posts from the three popular parenting Subreddits (i.e., topical communities on Reddit) r/Daddit, r/Mommit, and r/Parenting. These three parenting Subreddits gather different audiences, respectively, self-identifying as fathers and mothers (ostensibly single-gender), and parents (explicitly mixed-gender). By selecting a sample of users who have published on both a single-gender and a mixed-gender Subreddit, we are able to explore both audience and gender effects. We analyse posts with word embeddings by adding the username as a token in the corpus. This way, we are able to compare user-tokens to word-tokens and measure their similarity. We also investigate individual differences in this context by comparing users who exhibit significant changes in their behaviour (high self-monitors) with those who show less variation (low self-monitors). Results show that mothers and fathers, when interacting in a mixed-gender context, mostly behave similarly and discuss a great diversity of topics while fathers focus more on advising others on educational and family matters. In single-gender Subreddits, mothers and fathers are more focused on specific topics. Mothers in r/Mommit distinguish themselves from other groups by primarily discussing topics such as medical care, sleep and potty training, and food. Both mothers and fathers celebrate parenting events and describe or comment on the physical appearance of their children with a single-gender audience. In terms of individual differences, we find that, especially on r/Parenting, high self-monitors tend to conform more to the norms of the Subreddit by discussing more of the topics associated with the Subreddit. In conclusion, this study shows how mothers and fathers express different concerns and change their behaviour for different group-based audiences. It also advocates for the use of Reddit and word embeddings to better understand the dynamics of audience and gender in a natural setting.

\end{abstract}




\end{abstractbox}
%

\end{frontmatter}



\section*{Introduction}

Individuals identify with multiple social groups but only certain social identities are contextually salient. The way we behave in a given context depends on the audience. The effect of an audience on speakers' behaviour has been extensively documented since Goffman's concept of presentation of self or identity management \cite{goffman_presentation_1959}. He uses the metaphor of a theatre with a backstage and a front stage, arguing that individuals are like actors when interacting, performing to present a desirable image of themselves \cite{goffman_presentation_1959}. Individuals are very cautious about conforming to norms on the front stage, whereas they prepare their presentation backstage \cite{goffman_presentation_1959}. The social identity performance \cite{klein_social_2007} shows how the audience acts as a contextual identity clue, making social identity salient and therefore making relevant and appropriate a set of behaviours \cite{klein_social_2007}. These behaviours would position individuals in relation to the social identity in question in a way that allows them to achieve their social objectives \cite{klein_social_2007}. In addition to influencing individual behaviour, the audience's feedback shapes the definition of social identity too \cite{klein_social_2007}. In contrast to the concept of self-presentation, which focuses on being favourably seen as a personal goal, social identity performance considers the situations where the actor defines themselves as a group member and performs as such when they think they are visible to an audience \cite{klein_social_2007}. In 1974, Snyder extended Goffman's theory by taking an individual-level perspective, the Self-monitoring theory \cite{snyder1974self}. This theory holds that individuals differ in the extent to which they can and want to control their self-presentation. Specifically, individuals can be described as high or low self-monitors. High self-monitors are more careful about adapting their behaviour to the context they are in and matching the audience's expectations. On the other hand, low-self-monitors are less concerned with adapting their behaviour and conforming with social norms \cite{snyder1974self}.

The present work studies audience effects in parenting communities on Reddit. We use a novel method \cite{schuld_durrheim_mafunda_mazibuko_2022} of word embedding adding the username as a token in the text corpus. This method allows not only to locate agents from the way they talk but also to compare them to meaningful word-tokens. We use this method to focus on individuals and the influence gender and audience play in their interactions. To do so, we analyse a dataset of Reddit posts on the three popular parenting Subreddits r/Parenting, r/Daddit, and r/Mommit. We select a sample of Reddit users who have published on these Subreddits and study how the audience, either single-gender or mixed-gender, can influence the expression of concerns related to parenting.

Given the salience of gender stereotypes in the parenting domain \cite{wille19951990s, jolly2014gender, cinamon2002gender}, the gender configuration of the audience can push users towards specific behaviours, in conformity or not with social expectations regarding gender norms \cite{cinamon2002gender}. The salience of social identities related to gender can facilitate gender-stereotypical behaviours \cite{tajfel_social_1974, tajfel_integrative_1979, turner_rediscovering_1987, ellemers2018gender}. According to gender stereotypes, women have communal characteristics (e.g., care and selflessness) while men have agentic characteristics (e.g., motivation to master and assertiveness) \cite{eagly_gender_1984}. In practice, these stereotypes are reflected in gender roles as standards of `good' behaviour for men and women \cite{villicana_gender_2017}. These roles are associated with fields of expected expertise such as work for men and family for women \cite{ellemers2018gender}. For instance, although fathers want to be more involved in caring \cite{bulanda2004paternal}, they are positioned as "helpers" rather than as primary caregivers \cite{cancian2000caring, wall2007involved}.

Audience and gender stereotypes are therefore frameworks that we will use to understand behaviours. Do mothers and fathers adapt their expression of concerns to the audience they interact with? On which topics specifically? Is one gender more likely to change behaviour? How do users differ in their adaptation to audiences? These are questions we will address in the present study.

Reddit has several very popular parenting communities that make possible the study of those phenomena in a natural setting and at a large scale. We will use Reddit as a means through which we can observe how the change in audience can influence behaviour. Switching between different topical communities (i.e., Subreddits) and potentially social identities, with either a single or mixed-gender audience, mothers and fathers may adapt their behaviour accordingly. On Reddit, users' interactions on parenting have been described through several studies. Using a quantitative approach, Ammari et al. \cite{ammari_pseudonymous_2018} looked at the three popular parenting Subreddits r/Parenting, r/Mommit, and r/Daddit with Latent Dirichlet Allocation topic modelling \cite{blei_latent_2003}. They collected all the comments published on these Subreddits between 2008 and 2016, from which they were able to identify 47 topics. The authors found that users on r/Parenting talk more about discipline and teen talk. r/Mommit users most frequently discussed topics such as sleep training, breastfeeding, milestones, child weight gain, pregnancy recovery, and housework. On r/Daddit, the most frequent topics were congratulations, Neonatal Intensive Care Unit (NICU) experience, legal questions for custody battles, and Halloween costumes. Therefore, mothers discussed more issues related to health and raising children whereas fathers' discussions focused more on events such as birth, Halloween, or divorce. 

Using qualitative grounded theory, Feldman \cite{feldman__2021} analyzed posts published on r/Daddit and r/Mommit. They found that while fathers merge work and family and consider work as part of family duties, mothers see work and family as two distinct domains and emphasise the double burden. Mothers' posts are also more centred around \textit{how} to care for a child whereas fathers' are about \textit{whether} they care \cite{feldman__2021}. Additionally, in family relationships, fathers seek advice about how to provide support for the mother but mothers focus on a dyadic mother-child relationship \cite{feldman__2021}.

Sepahpour-Fard and Quayle \cite{sepahpour2022how} showed that the audience might play a role in the interactions parents have with each other. They selected a set of Reddit users who have published on both a single-gender Subreddit (r/Mommit or r/Daddit) and a mixed-gender Subreddit (r/Parenting) and looked at how users change their expression of concerns depending on the audience. They assumed, based on the Subreddits' descriptions, that r/Mommit is for mothers, r/Daddit for fathers, and r/Parenting for parents. Using Latent Dirichlet Allocation Topic Modelling \cite{blei_latent_2003}, they measured the prevalence of topics in comments and averaged the prevalence by gender and audience to isolate potential effects. Their results showed that mothers generally express more concerns about the children's basic needs (sleep, food, and medical care) than fathers. Interestingly, they express those concerns more between themselves (i.e., in r/Mommit) than with fathers (i.e., in r/Parenting). In contrast, fathers focus on giving educational advice in r/Parenting while celebrating events (e.g., pregnancies and births) and sharing pictures with other fathers in r/Daddit. In addition to the conformity of Reddit parents with gender roles regarding parenting, Sepahpour-Fard and Quayle \cite{sepahpour2022how} add a contextual approach which allows a better understanding of how users behave with different audiences. However, results are aggregated by groups of users and are therefore not showing what happens at the individual level. For example, users who publish more frequently will have a larger weight in the model and results might be showing what they are concerned with, rather than what all parents express. Moreover, some users might change their behaviour while others might discuss the same topics in different contexts.

In the present study, instead of looking at averages of posts similar to the previous studies, we use word embeddings (see Methods\ref{Methods}) and add the username as a token in the text to locate each user from the way they speak \cite{schuld_durrheim_mafunda_mazibuko_2022}. This way, we can give each user the same weight and analyse the expression of concerns at the individual level in parenting Subreddits. From this embedding where users are located, forming a speaker landscape \cite{schuld_durrheim_mafunda_mazibuko_2022}, we can see which topics users discuss, how focused they are (i.e., discussing only specific topics), and how they differently change behaviour interacting with different audiences. 

From comments published in 2020 on the three most popular parenting Subreddits (i.e. r/Daddit, r/Mommit, and r/Parenting), we selected only users who have published on both a single-gender Subreddit and a mixed-gender Subreddit, identifying mothers (i.e., users who publish on r/Mommit and r/Parenting) and fathers (users who publish on r/Daddit and r/Parenting). Their gender is inferred based on the Subreddits' descriptions (see Methods\ref{Methods} for more detail). This setting allows us to capture both gender differences and audience effects. We can see if mothers and fathers differently adjust their social identity performance when interacting with different audiences and if the location of users in the embedding shows significant insight regarding the change of behaviour.

\section*{Methods}
\label{Methods}
\subsection*{Data}

We collect comments published in 2020 on r/Daddit, r/Mommit, and r/Parenting, three communities on Reddit. Reddit is an American content aggregation platform. Users from any country can participate although English-speaking countries are largely over-represented \cite{alexa, similarweb}. Pseudonymous users publish content (e.g., links, text, images, or videos) on topical communities or `Subreddits' which are defined by descriptions and community rules, and governed by volunteer moderators. Subreddits can gather users around a large variety of topics such as politics, science, or video games. The broad theme of parenting has three popular Subreddits: r/Parenting with 4,431,185 members, r/Mommit with 538,017 members, and r/Daddit with 389,765 members as of June 2022.

r/Parenting is a mixed-gender Subreddit. It was created in April 2008, and its full name is ``Reddit Parenting - For those with kids of any age!". It is described as ``the place to discuss the ins and out as well as ups and downs of child-rearing. From the early stages of pregnancy to when your teenagers are finally ready to leave the nest (even if they don't want to), we're here to help you through this crazy thing called parenting. You can get advice on potty training, talk about breastfeeding, discuss how to get your baby to sleep or ask if that one weird thing your kid does is normal." 

r/Mommit is a mother-centric single-gender Subreddit. It was created in September 2010, and its full name is ``Mommit - Come for the support, stay for the details.". It is described as ``We are people. Mucking through the ickier parts of child-raising. It may not always be pretty, fun and awesome, but we do it. And we want to be here for others who are going through the same experiences and offer a helping hand." r/Mommit explicitly indicates that posters should be mothers.

r/Daddit is a father-centric single-gender Subreddit. It was created in August 2010. It is described as ``This is a Subreddit for Dads. Single Dads, new Dads, Step-Dads, tall Dads, short Dads, and any other kind of Dad. If you've got kids in your life that you love and provide for, come join us as we discuss everything from birth announcements to code browns in the shower.". Similarly to r/Mommit, r/Daddit is explicitly described to be exclusively for fathers.

We collected comments using Pushshift Reddit Application Programming Interface (API)\footnote{Code in Python by Mikołaj Biesaga, The Robert Zajonc Institute for Social Studies, University of Warsaw. GitHub repository: https://github.com/MikoBie/reddit} \cite{baumgartner2020pushshift}. Pushshift is a platform collecting social media data in real-time. Then, we selected only users who have published on both a single-gender Subreddit (r/Mommit or r/Daddit) and a mixed-gender Subreddit (r/Parenting), on the assumption that posting on "mommit" or "daddit" indicates identification as a mother or father, in accordance with the Subreddit descriptions. Before further preprocessing, our dataset contains 8,361 unique Reddit users who published 194,497 comments. 

\subsubsection*{Preprocessing}
We first remove from the dataset 502 authors who have published comments both on r/Mommit and r/Daddit, as we cannot know whether they primarily identify as ``mother" or ``father". We remove comments written by the auto-moderator and deleted comments. We tokenize the documents, create ngrams, and remove a list of stop words, i.e., words occurring in abundance but providing little or no information about the content of text documents, using Scikit-learn \cite{pedregosa2011scikit} to which we add ``ai", ``im", ``m", ``s", ``ve", ``w", ``d", ``ive", ``id", ``itll", ``shes", ``hes", ``theyre", ``youre", ``dont", ``got", ``havent", ``didn". We finally lemmatize the text tokens and remove comments that became empty because of the cleaning process. The dataset now includes 148,304 posts with 24,035 posts on r/Mommit, 16,388 posts on r/Daddit, 78,450 posts on r/Parenting by mothers, and 29,431 posts on r/Parenting by fathers.

Before running the word embedding algorithm, we added the username as a token in the text. The position of the agent token was centred, i.e., for each comment, we take its length (i.e., number of words) and the user-token is added at the centre of the comment. We believe this position to be the best in order to have an appropriate representation of users taking into account the surrounding word-tokens.

\subsection*{Word embeddings}
\subsubsection*{Definition}

A word embedding is a Natural Language Processing technique representing words as vectors. Words which share common linguistic characteristics (e.g., semantics and syntax) are located close to each other spatially. In other words, if two words having the same context (i.e., the same surrounding words) are passed as inputs, the outputs for these two words should be similar.

To create the embedding, we use Word2Vec in the Gensim library \cite{rehurek_lrec}. Word2Vec is one of the algorithms which produces vectors for words in textual documents. It can be used through two models: the Continuous Bag-Of-Words model or the Skip-Gram model. While the Continuous Bag-Of-Words model predicts a target word from its context (i.e., surrounding words), the Skip-Gram model predicts the context from target words and is better for larger datasets. Parameters include the choice of model (i.e., Continuous Bag-Of-Words or Skip-Gram), subsampling (e.g., removing too frequent or rare words), dimensionality (i.e., the size of the vector), and window size (i.e., number of tokens to be considered as the context of the target word). To train the word embedding algorithm, we used the Skip-Gram model, removed infrequent tokens occurring less than five times in the corpus, set the size of the vector to 200 dimensions, and a window of 15 tokens, as 77\% of posts are less than 30-tokens long (a window of 15 means the algorithm will look at the 15 words before and the 15 words after the target word). As a robustness check, we ran the model with three other random seeds and different vector sizes (100, 150, and 200) and found consistent results.

\subsubsection*{User embeddings and the concept of speaker landscapes}
User embeddings represent users as vectors in an embedding space. By spatially projecting users, we can analyze the relationship between users, their concentration or spread in the embedding, and group dynamics. User embeddings can be used to detect sarcasm \cite{amir2016modelling}, build recommendations systems \cite{pal2020pinnersage, yu-etal-2016-user}, predict gender and depression \cite{xiaodong2020author}, etc. In the present study, we create a `speaker landscape', a novel method developed by Schuld et al. \cite{schuld_durrheim_mafunda_mazibuko_2022} where the username is added in the text document as a token. As word embeddings create a vector representation for a target word from its surrounding words, the target word vector represents its context. When adding the user as a token in a document, the embedding algorithm will locate the user spatially relying on the words used in the document. To create a `speaker landscape' \cite{schuld_durrheim_mafunda_mazibuko_2022}, we add the username at the central position of the Reddit post they published. This method allows to create a vector representation of users from the words they use and, as the user is a word-token in the text, it can be directly compared to other word-tokens (e.g. how close r/Daddit fathers are to words such as `food', `sleep', or `health'?).  

\subsubsection*{Visualizing the embedding}
As the output of the embedding algorithm gives us 200-dimensional vectors, we need to reduce the vectors into two-dimension vectors to be able to visualise them. For reducing vectors, we use Uniform Manifold Approximation and Projection (UMAP) \cite{https://doi.org/10.48550/arxiv.1802.03426}. UMAP uses graph layout algorithms to represent data as a high-dimensional graph from which it builds a low-dimensional graph with a similar structure. Two important parameters, \texttt{n\_neighbors} and \texttt{min\_dist}, are used to control the balance between local and global structure. A low value of \texttt{n\_neighbors} will make UMAP consider more the local structure, focusing on the closest points to analyse the high-dimensional data whereas a high value will represent more the bigger picture. \texttt{min\_dist} corresponds to the minimal distance between the points represented in low-dimensional space, i.e., low values will make the embeddings appear tightly packed together. It should be noted that, as UMAP has to warp the high-dimensional space to project the data to lower dimensions, the axes or distances in the low-dimensional representation are not directly interpretable.

For visualising our results, we keep users who posted at least five comments on each Subreddit to have a more robust position of users in the word embedding. This leaves us 2872 users to be visualized, 956 mothers and 480 fathers. Each user will be represented by two dots, each for an audience with which they interact (either single-gender or mixed-gender).

Additionally, as users are tokens along with other word-tokens in our corpus, we can directly compare their position with the position of other words. To have a meaningful set of word-tokens, we use the results of topic modelling from a previous study conducted on the same dataset \cite{sepahpour2022how}. \cite{sepahpour2022how} found twelve topics to best represent discussions in r/Daddit, r/Mommit, and r/Parenting Subreddits: \textit{Thank you/Appreciation} (i.e., thanking other users for their advice), \textit{Medical care}, \textit{Education/Family advice} (i.e., giving advice to other parents), \textit{Furniture/Design} (i.e., discussing furniture for children in the house), \textit{Birth/Pregnancy} (i.e., announcements or congratulations related to child arrival), \textit{Change/Potty training}, \textit{Physical appearance/Picture} (i.e., commenting on the physical appearance of children and sharing pictures), \textit{Work/Raise children} (i.e., discussing the balance between work and family), \textit{Food}, \textit{Leisure activities}, \textit{School/Teaching}, and \textit{Sleep training}. For each topic, we select six keywords among the topics' ten most associated keywords and add them to the visualisation.

In order to look at the differences between high and low self-monitors \cite{snyder1974self}, we analysed the distribution of cosine similarity between two versions of the same user (on a single-gender Subreddit and on a mixed-gender Subreddit). We defined low self-monitors as the 25\% of the users with the largest similarity scores and the high self-monitors as the 25\% of the users with the smallest similarity scores.

\subsubsection*{Measures used to analyze the output}
To measure the relationship between two tokens, we measure the similarity between their vector representation. For this, we use cosine similarity (i.e., the dot product of the vectors divided by the product of their length), which is defined as follows:

\begin{equation}
\cos (\theta)= {\frac{\mathbf {A} \cdot \mathbf {B}} {\|\mathbf {A} \|\|\mathbf {B} \|}} = \frac{ \sum_{i=1}^{n}{\mathbf {A}_i\mathbf {B}_i} }{ \sqrt{\sum_{i=1}^{n}{\mathbf {A}_i^2}} \sqrt{\sum_{i=1}^{n}{\mathbf {B}_i^2}} }
\end{equation}

The cosine values are between -1 and 1 and the greater the cosine value, the more similar the two tokens are. To compare the similarity measures between two groups of users or the same users but in two different contexts, we used the Kolmogorov-Smirnov test \cite{10.2307/2280095} and Student's t-test \cite{10.1093/biomet/6.1.1}. The Kolmogorov-Smirnov test for two samples gives the probability that the two samples were drawn from the same distribution. The Student's t-test is used to compare the means of two samples.

\section*{Results}
\subsection*{Speaker landscape}
\begin{figure}[ht!]
    \includegraphics[width=0.98\linewidth]{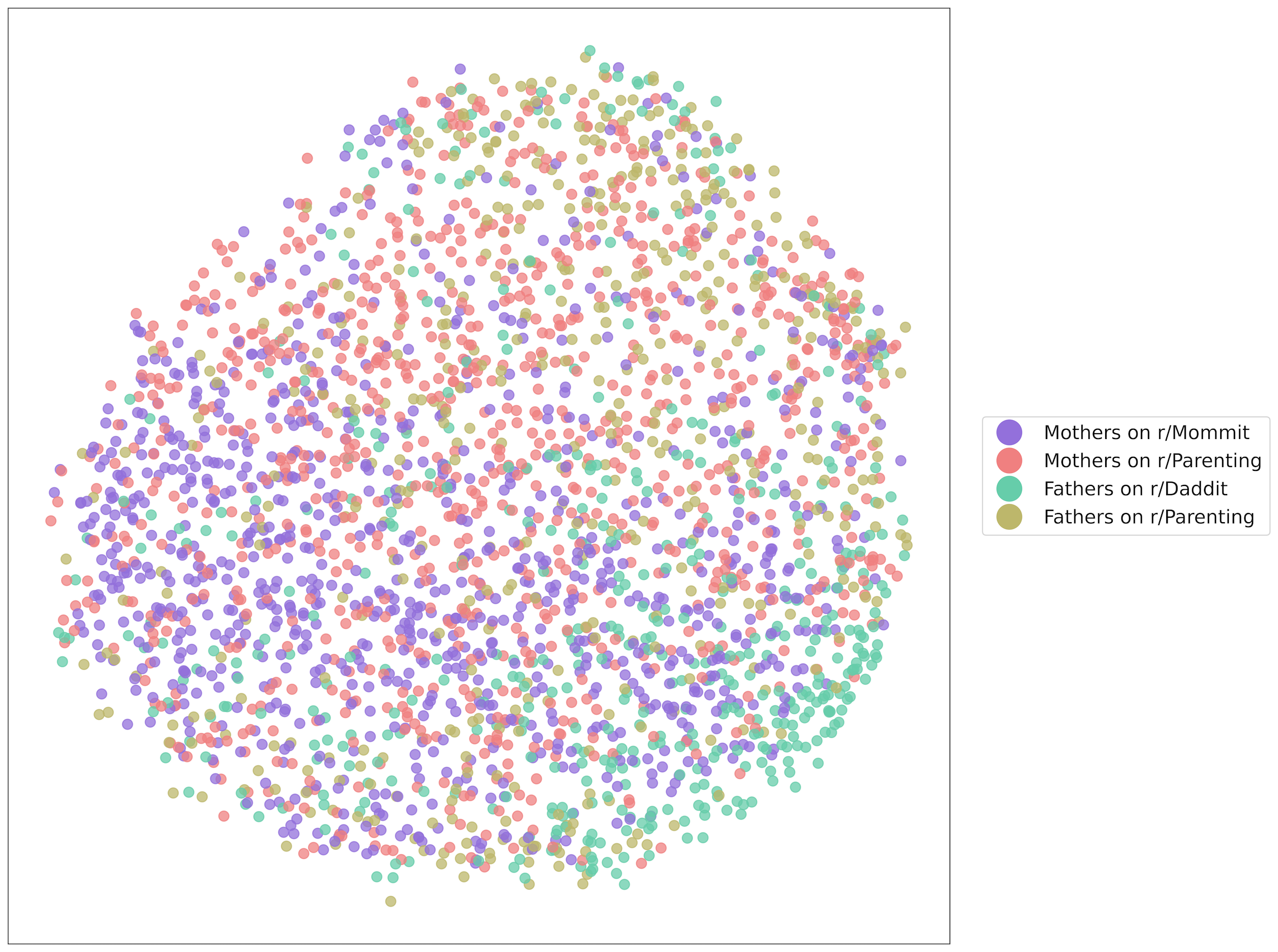}
    \caption{\csentence{Speaker landscape of mothers and fathers interacting with different audiences.}
  Each dot represents an author posting on one Subreddit (either r/Daddit, r/Mommit, or r/Parenting).}
      \label{figure1}
\end{figure}

We use the term `speaker landscape' to refer to the user embedding created by adding the username as a token in the text. We begin by visualising the vector representation of Reddit users and qualitatively study the structure of the landscape. In Figure \ref{figure1}, we can see the two-dimensional representation of users coloured by their self-identified status as mothers or fathers, according to the Subreddit on which they posted. The two-dimensional representation of the usernames' vectors was obtained using UMAP dimension reduction technique \cite{https://doi.org/10.48550/arxiv.1802.03426}. Each user has two data points representing them, one when interacting with a single-gender audience (posting on r/Mommit or r/Daddit) and one when interacting with a mixed-gender audience (posting on r/Parenting). 

Fathers on r/Daddit are clustered at the right of the landscape. In contrast, mothers on r/Mommit are mostly on the bottom half of the landscape and are more spread out. In the presence of a mixed-gender audience (r/Parenting), both fathers and mothers seem to discuss a wide range of topics as their data points are spread out around the landscape.

To quantify how spatially concentrated or dispersed each group is in the embedding space, we measure the average cosine similarity between the vector representations of each pair of users in a group (mothers on r/Mommit, mothers on r/Parenting, fathers on r/Daddit, and fathers on r/Parenting). We then measure the differences between the concentration of groups using T-tests and Kolmogorov-Smirnov tests (see Methods\ref{Methods} for more detail). Results show that all the differences between the different groups are significant (see Additional Files\ref{addfiles}, Table \ref{table2}). Fathers, when interacting with other fathers are the most concentrated group, meaning they tend to gather more around specific topics than other groups. Mothers in r/Mommit are also more concentrated than users on r/Parenting. When in a mixed-gender context on r/Parenting, parents interact about a greater set of topics than when in single-gender contexts and are therefore less spatially concentrated. Looking at mothers and fathers within the context of r/Parenting, Mothers are the most spatially spread group. These results confirm the qualitative analysis of the speaker landscape with fathers in r/Daddit being the most concentrated group followed by mothers in r/Mommit, fathers in r/Parenting, and mothers in r/Parenting (Figure \ref{figure1}).

Figure \ref{figure2} shows the probability density function of the cosine similarity. In addition to all differences appearing to be significant (see Additional Files\ref{addfiles}, Table \ref{table2}), we can see two distinct types of behaviour among both mothers and fathers: they are more focused on specific topics of discussion when interacting with a single-gender audience while they diversify topics when gathered in a mixed-gender environment. 

\begin{figure}[ht!]
    \includegraphics[width=0.98\linewidth]{Figure2_similarity_within_groups.png}
  \caption{\csentence{Distribution of similarity between users within each group of gender and audience.}
      The higher the cosine similarity value, the more similar agents are to each other.}
      \label{figure2}
      \end{figure}

\subsection*{Similarity between two versions of the same author}
\label{2versions_similarity}
Are there gender differences between mothers and fathers? i.e., Does one gender change their behaviour more when switching from one Subreddit to the other? We use the cosine similarity between two vectors (associated with the two data points representing each user in the two Subreddits) to measure the extent to which users change their discourse when interacting with different audiences (Figure \ref{figure3}). Figure \ref{figure3} shows the distributions of cosine similarity values between two versions of the same user, i.e., the cosine similarity between the same user interacting on r/Mommit or r/Daddit (not both) and interacting on r/Parenting. The Kolmogorov-Smirnov test comparing two distributions shows that the difference between mothers and fathers is not significant. Therefore, as there is no evidence of difference, there might be a similar tendency for mothers and fathers to be more focused on specific topics within gender-specified contexts, but more diverse when gathered in the more pluralistic r/Parenting Subreddit.

\begin{figure}[ht!]
\includegraphics[width=0.98\linewidth]{Figure3_similarity_2versions_parents_distrib.png}
  \caption{\csentence{Distribution of similarity between two versions of the same user for mothers and fathers}
      The higher the cosine similarity, the less the user changes behaviour between Subreddits.}
      \label{figure3}
      \end{figure}

\subsection*{Topics of discussion}
Previous work by Sepahpour-Fard and Quayle \cite{sepahpour2022how} found, using Latent Dirichlet Allocation Topic Modelling, twelve topics in users' discourse on r/Mommit, r/Daddit, and r/Parenting: \textit{Food}, \textit{Sleep training}, \textit{Medical care}, \textit{Change/Potty training}, \textit{Birth/Pregnancy}, \textit{Thank you/Appreciation}, \textit{Physical appearance/Picture}, (\textit{Education/Family advice}, \textit{Work/Raise children}, \textit{School/Teaching}, \textit{Leisure activities}, and \textit{Furniture/Design}. Each topic was associated with ten keywords representing them (more details available in \cite{sepahpour2022how}).
To validate these keywords and topics, we first visualise the topic modelling keywords found in this study \cite{sepahpour2022how} (Figure \ref{figure4}). The keywords are coloured by the topic they are the most associated with. As expected, and validating our speaker landscape, each topic's keywords are clustered together in the embedding.

\begin{figure}[ht!]
    \includegraphics[width=0.98\linewidth]{Figure4_topics_keywords.png}
  \caption{\csentence{Topic Modelling keywords} Each colour represents a topic found through LDA topic modelling on the same dataset.}
  \label{figure4}
      \end{figure}

To identify which topics are the most discussed by each group of users, we measure the distance between topics and users interacting in the different Subreddits by measuring the cosine similarity between keywords in a topic and usernames in each group (Figure \ref{figure5}). First, results show the influence of Subreddits' norms in r/Parenting, as the topics \textit{Thank you/Appreciation}, \textit{Medical care}, \textit{Birth/Pregnancy}, \textit{Change/Potty training}, \textit{Physical appearance/Picture}, and \textit{Sleep training} are similarly discussed by mothers and fathers on the Subreddit. These same topics differ in the way they are discussed in single-gender Subreddits with fathers on r/Daddit discussing \textit{Thank you/Appreciation}, \textit{Birth/Pregnancy}, and \textit{Physical appearance/Picture} the most and mothers on r/Mommit discussing \textit{Medical care}, \textit{Change/Potty training}, and \textit{Sleep training} the most. The topic \textit{Furniture/Design}, mainly discussed in r/Mommit and r/Daddit, seems to distinguish single-gender Subreddits and mixed-gender Subreddits.  The topics \textit{Education/Family advice} and \textit{School/Teaching} are mostly discussed by fathers in r/Parenting, therefore showing gender differences in the same Subreddit. The topic \textit{Leisure activities} is discussed by fathers in r/Daddit the most and by mothers in r/Mommit the least. The topic \textit{Work/Raise children} is mostly discussed by fathers, especially on r/Daddit. Finally, the topic \textit{Food} clearly distinguishes mothers in r/Mommit from other groups.

\begin{figure}[ht!]
    \includegraphics[width=0.98\linewidth]{Figure5_grouped_boxplot_allgroups.png}
  \caption{\csentence{Variability of cosine similarity across topics and groups}
      The higher the cosine similarity, the more the agent is focused on that topic.}
      \label{figure5}
      \end{figure}

We further analyse individual differences by focusing on the one hand, users who change their behaviour a lot, and on the other hand, those who behave similarly across audiences. We analyse each topic and see how the results gathering all authors can be compared with the results for high and low self-monitoring mothers (Figure \ref{figure6}) and fathers (Figure \ref{figure7}). We expect the high self-monitors (i.e., the least similar authors) to be closer to topics highly associated with a specific context, e.g., if Physical appearance/Picture is highly associated with the single-gender Subreddits, we would expect high self-monitors to be more similar with this topic in these Subreddits than low self-monitors. High self-monitoring fathers adapt themselves when interacting on r/Parenting. For instance, the topics \textit{Education/Family advice} and \textit{School/Teaching}, which were mostly discussed by fathers on r/Parenting, are even more discussed by high self-monitoring fathers. Similarly, as fathers discussed \textit{Furniture/Design}, \textit{Work/Raise children}, \textit{Food}, and \textit{Sleep training} more in r/Daddit than in r/Parenting, high self-monitoring fathers discuss the topics the most. High self-monitoring fathers on r/Parenting are also the ones discussing the least the topics more associated with r/Daddit than with r/Parenting (\textit{Thank you/Appreciation}, \textit{Medical care}, \textit{Furniture/Design}, \textit{Birth/Pregnancy}, \textit{Change/Potty training}, \textit{Physical appearance/Picture}, and \textit{Sleep training}). Low self-monitoring fathers in r/Parenting, are never the ones discussing topics the most. In contrast, low self-monitoring fathers in r/Daddit discuss \textit{Thank you/Appreciation}, \textit{Birth/Pregnancy}, and \textit{Physical appearance/Picture} the most. We observe similar results for mothers. When the topics are more associated with r/Mommit than r/Parenting, high self-monitoring mothers are discussing these topics on r/Parenting the least (\textit{Thank you/Appreciation}, \textit{Medical care}, \textit{Furniture/Design}, \textit{Birth/Pregnancy}, \textit{Change/Potty training}, \textit{Physical appearance/Picture}, \textit{Food}, \textit{Leisure activities}, and \textit{Sleep training}) while they discuss topics associated with r/Parenting the most (\textit{Education/Family advice} and \textit{School/Teaching}). Similarly to fathers, low self-monitoring mothers in r/Mommit discuss \textit{Thank you/Appreciation}, \textit{Birth/Pregnancy}, \textit{Physical appearance/Picture}, and \textit{Sleep training} the most, which are associated with single-gender Subreddits.

\begin{figure}[ht!]
    \includegraphics[width=0.98\linewidth]{Figure6_grouped_boxplot_mothers.png}
  \caption{\csentence{Variability of cosine similarity across topics and groups - High and low self-monitoring mothers}
      The higher the cosine similarity, the more the agent is focused on that topic.}
      \label{figure6}
      \end{figure}

\begin{figure}[ht!]
    \includegraphics[width=0.98\linewidth]{Figure7_grouped_boxplot_fathers.png}
  \caption{\csentence{Variability of cosine similarity across topics and groups - High and low self-monitoring fathers}
      The higher the cosine similarity, the more the agent is focused on that topic.}
      \label{figure7}
      \end{figure}

\section*{Discussion}
We explored the effects of audience and gender on interactions in parenting Subreddits and suggest a new method to measure self-monitoring in individuals. We used the speaker landscape as a method for the study of individuals, groups, and group interactions. The results are consistent with previous results, which analyzed the same dataset using topic modelling \cite{sepahpour2022how} and therefore gave us additional confidence about the position of usernames. However, going beyond topic modelling results, the speaker landscape approach could capture how individual users change their behaviour when switching Subreddits without having to rely on averages biased towards those who post a lot. 

Coherently with previous works on the effect of an audience on behaviour \cite{klein_social_2007, goffman_presentation_1959, turner_rediscovering_1987}, we found that both mothers and fathers change their topics of discussion when interacting with different audiences. We looked at the differences between groups (i.e., mothers on r/Mommit, fathers on r/Daddit, mothers on r/Parenting, and fathers on r/Parenting), and saw two clusters of behaviour appearing with users in a single-gender Subreddit being more focused (i.e., higher cosine similarity value) than users in a mixed-gender Subreddit. Results also showed that parents, when interacting in a mixed-gender Subreddit, were more diverse linguistically. In contrast, in a single-gender Subreddit, mothers and fathers were more clustered in the landscape which indicates fewer topics of discussion. Previous research \cite{hilte2022linguistic} showed that teenage girls and boys (aged 13 to 20) tend to converge to more similar behaviour when in a mixed-gender group, girls and boys adopt less gender-specific writing styles. We observe the same behaviour, with mothers and fathers tending to be less `specialized' in specific topics (more spread in the landscape) when gathered in a mixed-gender Subreddit. This phenomenon is also explained by the Communication Accommodation Theory \cite{giles1979accommodation}, which describes how individuals adjust their language to each other to get approval, communicate more efficiently, and maintain a positive social identity. The latter might explain why users differ more when in mixed-gender groups than when gathered in a single-gender group. However, at an even more abstract level, groups and social identities, are founded in language; and language is a primary way for people to position themselves as group members (especially in online contexts where it is the only medium for social interaction and group definition).

We found that the topics stereotypically associated with mothers such as food, potty training, or sleep training appear only when they post in a single-gender Subreddit. These are topics particularly associated with the gender role of mothers \cite{ellemers2018gender} as they are about the basic needs of children and the practical aspects of raising a child. By discussing them the most and only when between themselves, they position themselves as more concerned by these areas. In a single-gender Subreddit, fathers discussed more topics related to announcing births and pregnancies, appreciation, shared pictures and commented on physical appearance. We also found, thanks to our method which separates mothers and fathers, that behaviours are not homogeneous in a Subreddit. For instance, in r/Parenting, fathers seem to be particularly concerned with education, giving advice and discussing school and teaching activities, distinguishing themselves from mothers on the same Subreddit. Previous studies on the same parenting Subreddits \cite{ammari_pseudonymous_2018, feldman__2021, sepahpour2022how} found similar results with users discussing more sleep training, breastfeeding, and potty training on r/Mommit, congratulations on r/Daddit and, discipline on r/Parenting.

Extending previous work using those topics and taking advantage of the speaker landscape method that allows us to have a more individual-level approach, we further analysed two groups of users: high and low self-monitoring parents \cite{snyder1974self}. With the speaker landscape, we were able to measure the extent of behaviour change in users. This way, we could identify low self-monitors (users maintaining constant behaviour across contexts) and high self-monitors (users adapting their behaviours across contexts). Results were mostly consistent with the self-monitoring theory \cite{snyder1974self}, with high self-monitors changing more behaviour across contexts than low self-monitors. However, we unexpectedly found that both low self-monitoring fathers and mothers in single-gender Subreddits were discussing more topics associated with these Subreddits. These results might suggest that these authors specialized in those topics and focused on those when interacting in single-gender Subreddits.


Although allowing a deep analysis of parents' interactions on Reddit, our analysis has some limitations related to the method used. Future work should focus on untangling the potential biases caused by the method. As word embeddings are initially made for representing words as vectors and highlighting relationships between words, adding a non-word-token (i.e., the user-token) might misrepresent other word-tokens. For instance, an important part of the word-tokens is at the periphery of the landscape. This might be explained by the fact that ngrams surrounding user-tokens might be more diverse than those surrounding other word-tokens, therefore making user-tokens more central in the landscape. In other words, if the word "bed" is most of the time surrounded by words such as "night" or "sleep", a user-token can once be surrounded by words such as "night" or "sleep" when talking about sleep training but by "school" and "teaching" when discussing education. More generally, the speaker landscape distorts language by adding a non-word token in the text. The consequences of the latter should be studied to prevent biases in the representations of users and words. In our work, however, the speaker landscape is validated by the clustered topics' keywords in the landscape and the results are coherent with previous studies on the same Subreddits. 

The theories, dataset and topics we used in the present study might not represent globally applicable gender norms and parenting behaviours. For instance, Reddit is mainly used in English-speaking countries \cite{amaya2021new}, thus its data and our conclusions might not represent parents in all cultures. One should therefore be careful before assuming such results apply to any cultural context.

Additionally, although we improved previous research by looking at individual differences and giving the same weight to each user, we stayed focused on the topics found in a previous study \cite{sepahpour2022how}. There might be differences that are not captured by these topics and are more linguistically subtle. Future research should look deeper into linguistic features such as linguistic style variation.
Using word embeddings, we represented users spatially to analyse audience effects in parenting Subreddits and individual differences. Our results show that mothers and fathers focus on specific topics when in r/Mommit and r/Daddit but converge into more similar topics when gathered in r/Parenting. In a single-gender context, mothers preferentially discuss matters regarding the basic needs of children (food, sleep, health) while fathers interact the most about physical appearance and photos of children. We explored how low and high self-monitors are differently impacted by the change in audience. Our study demonstrates how Reddit can be used as a means through which researchers can study audience effects and behaviours at a large scale while taking into account individual behaviour.


\begin{backmatter}
\section*{Declarations}
\subsection*{Availability of data and materials}
The datasets generated and analysed during the current study are not publicly available due to Reddit's Terms and Conditions of use but are available from the corresponding author on reasonable request.

\subsection*{Competing interests}
  The authors declare that they have no competing interests.

\subsection*{Funding}
This publication has emanated from research conducted with the financial support of Science Foundation Ireland under Grant number 18/CRT/6049 and the European Research Council (ERC) under the European Union's Horizon 2020 research and innovation programme (grant agreement No. 802421).

\subsection*{Author's contributions}
MSF collected the data, analysed the data, and wrote the paper.
MS provided the code for analysis.
MQ and TY conceptualised and designed the study.
All authors contributed to and approved the paper.


\bibliographystyle{bmc-mathphys} 
\bibliography{reddit_parenting}      





\section*{Additional files}
\label{addfiles}


\subsection*{Similarity between Reddit users}
Table \ref{table1} shows the average and standard deviation values of the cosine similarity between users in each group. The figure reproduces the pattern found in Figure 2 with r/Mommit and r/Daddit users being more tightly gathered in the landscape while r/Parenting users being less similar and therefore more spread out in the landscape.

Table \ref{table2} shows the test statistics and the significance of Kolmogorov-Smirnov tests conducted to compare the distribution of cosine similarity values between groups. All differences were significant. 

\hspace{2em}

\begin{center}
\begin{table}[!ht]
\begin{tabular}{ |l|c|c| }
\hline
Groups & Average similarity between agents & Standard Deviation \\ \hline
Fathers on r/Daddit & 0.303 & 0.109\\
\hline
Mothers on r/Mommit & 0.278 & 0.104\\
\hline
Fathers on r/Parenting & 0.204 & 0.088\\
\hline
Mothers on r/Parenting & 0.189 & 0.085\\
\hline

\end{tabular}
\caption{\label{tab:groups_similarity_with_centroid_average}Average similarity with centroid and standard deviation for each group.}
\label{table1}

\hspace{2em}

\begin{tabular}{ |l|c|c|c|c|c|}
\hline
 Groups & Fathers on r/Daddit & Mothers on r/Mommit & Fathers on r/Parenting & Mothers on r/Parenting \\ \hline
Fathers on r/Daddit & - & - & - & -\\
\hline
Mothers on r/Mommit & 0.094\mbox{***} & - & - & -\\
\hline
Fathers on r/Parenting & 0.390\mbox{***} & 0.310\mbox{***} & - & -\\
\hline
Mothers on r/Parenting & 0.450\mbox{***} & 0.373\mbox{***} & 0.068\mbox{***} & - \\
\hline

\end{tabular}
\caption{\label{tab:groups_similarity_with_centroid_kstest}Kolmogorov-Smirnov test statistics for the differences between groups of users. P-values less than 0.05 are summarized with one asterisk (*) and P-values less than 0.001 are summarized with three asterisks (***). No asterisk means that no significant difference was found.}
\label{table2}
\end{table}
\end{center}

\end{backmatter}
\end{document}